\documentclass{bmcart}

\usepackage[utf8]{inputenc} 
\usepackage{multirow}
\usepackage{caption}
\usepackage{siunitx}
\usepackage[numbers]{natbib}

\def\includegraphics{}

\usepackage{graphicx}

\startlocaldefs
\endlocaldefs

\begin{document}

\begin{frontmatter}

\begin{fmbox}
\dochead{Research}


\title{Feasibility study of the positronium lifetime imaging with the Biograph Vision Quadra and J-PET tomographs}


\author[
   addressref={fais,ct},                   
   email={szymon.parzych@doctoral.uj.edu.pl}   
]{\inits{SP}\fnm{Szymon} \snm{Parzych}}
\author[
   addressref={fais,ct}
]{\inits{SN}\fnm{Szymon} \snm{Niedźwiecki}}
\author[
   addressref={fais,ct}
]{\inits{EYB}\fnm{Ermias} \snm{Yitayew Beyene}}
\author[
   addressref={fais,ct}
]{\inits{NC}\fnm{Neha} \snm{Chug}}
\author[
   addressref={siemens}
]{\inits{MC}\fnm{Maurizio} \snm{Conti}}
\author[
   addressref={lnf}
]{\inits{CC}\fnm{Catalina} \snm{Curceanu}}
\author[
   addressref={fais,ct}
]{\inits{EC}\fnm{Eryk} \snm{Czerwiński}}
\author[
   addressref={fais,ct}
]{\inits{MD}\fnm{Manish} \snm{Das}}
\author[
   addressref={fais,ct}
]{\inits{KVE}\fnm{Kavya} \snm{Valsan Eliyan}}
\author[
   addressref={agh}
]{\inits{JH}\fnm{Jakub} \snm{Hajduga}}
\author[
   addressref={fais,ct}
]{\inits{SJ}\fnm{Sharareh} \snm{Jalali}}
\author[
   addressref={fais,ct}
]{\inits{KK}\fnm{Krzysztof} \snm{Kacprzak}}
\author[
   addressref={fais,ct}
]{\inits{TK}\fnm{Tevfik} \snm{Kaplanoglu}}
\author[
   addressref={fais,ct}
]{\inits{ŁK}\fnm{Łukasz} \snm{Kapłon}}
\author[
   addressref={fais,ct}
]{\inits{KK}\fnm{Kamila} \snm{Kasperska}}
\author[
   addressref={fais,ct}
]{\inits{AK}\fnm{Aleksander} \snm{Khreptak}}
\author[
   addressref={fais,ct}
]{\inits{GK}\fnm{Grzegorz} \snm{Korcyl}}
\author[
   addressref={fais,ct}
]{\inits{TK}\fnm{Tomasz} \snm{Kozik}}
\author[
   addressref={fais,ct}
]{\inits{DK}\fnm{Deepak} \snm{Kumar}}
\author[
   addressref={fais,ct}
]{\inits{AKV}\fnm{Anoop} \snm{Kunimmal Venadan}}
\author[
   addressref={fais,ct}
]{\inits{KK}\fnm{Karol} \snm{Kubat}}
\author[
   addressref={pk}
]{\inits{EL}\fnm{Edward} \snm{Lisowski}}
\author[
   addressref={pk}
]{\inits{FL}\fnm{Filip} \snm{Lisowski}}
\author[
   addressref={fais,ct}
]{\inits{JMS}\fnm{Justyna} \snm{Medrala-Sowa}}
\author[
   addressref={fais,ct}
]{\inits{WM}\fnm{Wiktor} \snm{Mryka}}
\author[
   addressref={fais,ct}
]{\inits{SM}\fnm{Simbarashe} \snm{Moyo}}
\author[
   addressref={fais,ct}
]{\inits{PP}\fnm{Piyush} \snm{Pandey}}
\author[
   addressref={fais,ct}
]{\inits{EP}\fnm{Elena} \snm{Perez del Rio}}
\author[
   addressref={fais}
]{\inits{AP}\fnm{Alessio} \snm{Porcelli}}
\author[
   addressref={agh}
]{\inits{BR}\fnm{Bartłomiej} \snm{Rachwał}}
\author[
   addressref={fais,ct}
]{\inits{MR}\fnm{Martin} \snm{Rädler}}
\author[
   addressref={bern}
]{\inits{AR}\fnm{Axel} \snm{Rominger}}
\author[
   addressref={fais,ct}
]{\inits{SS}\fnm{Sushil} \snm{Sharma}}
\author[
   addressref={bern}
]{\inits{KS}\fnm{Kuangyu} \snm{Shi}}
\author[
   addressref={fais,ct}
]{\inits{MS}\fnm{Magdalena} \snm{Skurzok}}
\author[
   addressref={siemens}
]{\inits{WMS}\fnm{William M.} \snm{Steinberger}}
\author[
   addressref={agh}
]{\inits{TS}\fnm{Tomasz} \snm{Szumlak}}
\author[
   addressref={fais,ct}
]{\inits{PT}\fnm{Pooja} \snm{Tanty}}
\author[
   addressref={fais,ct}
]{\inits{KTA}\fnm{Keyvan} \snm{Tayefi Ardebili}}
\author[
   addressref={fais,ct}
]{\inits{ST}\fnm{Satyam} \snm{Tiwari}}
\author[
   addressref={fais,ct}
]{\inits{EŁS}\fnm{Ewa Ł.} \snm{Stepień}}
\author[
   addressref={fais,ct}
]{\inits{PM}\fnm{Paweł} \snm{Moskal}}


\address[id=fais]{
  \orgname{Faculty of Physics, Astronomy and Applied Computer Science, Jagiellonian University}, 
  \street{ul. prof. Stanisława Łojasiewicza 11},                     %
  \postcode{30-348}                                
  \city{Kraków},                              
  \cny{Poland}                                    
}

\address[id=ct]{
  \orgname{Center for Theranostics, Jagiellonian University}, 
  \street{ul. Mikołaja Kopernika 40},                     %
  \postcode{31-034}                                
  \city{Kraków},                              
  \cny{Poland}                                    
}

\address[id=siemens]{
  \orgname{Siemens Medical Solutions USA, Inc., Knoxville, TN 37932}, 
  \cny{USA}                                    
}

\address[id=lnf]{
  \orgname{Laboratori Nazionali di Frascati, Istituto Nazionale di Fisica Nucleare}, 
  \street{Via Enrico Fermi 54},                     %
  \postcode{00044}                                
  \city{Frascati},                              
  \cny{Italy}                                    
}

\address[id=agh]{
  \orgname{AGH University of Kraków}, 
  \street{al. Adama Mickiewicza 30},                     %
  \postcode{30-059}                                
  \city{Kraków},                              
  \cny{Poland}                                    
}

\address[id=pk]{
  \orgname{Cracow University of Technology, Faculty of Mechanical Engineering}, 
  \postcode{31-864}                                
  \city{Kraków},                              
  \cny{Poland}                                    
}

\address[id=bern]{
  \orgname{Department of Nuclear Medicine, Inselspital, Bern University Hospital, University of Bern}, 
  \city{Bern},                              
  \cny{Switzerland}                                    
}


\begin{artnotes}
\note[id=n1]{...} 
\end{artnotes}

\end{fmbox}


\begin{abstractbox}

\begin{abstract} 
\parttitle{Background} 
After its first ex-vivo and in-vivo demonstration, Positronium Lifetime Imaging (PLI) has received considerable interest as a potential new diagnostic biomarker. High sensitivity Positron Emission Tomography (PET) systems are needed for PLI since it requires simultaneous registration of annihilation photons and prompt gamma. In this simulation-based study, a~feasibility of PLI with the long axial field-of-view Biograph Vision Quadra (Quadra) and the Total Body J-PET scanner was investigated.

\parttitle{Methods}
The study was performed using the GATE software. Background radiation, present within the Quadra tomograph, was added to the simulation. First, the optimal placement of the energy window for the registration of the prompt gamma was investigated. Next, the organ-wise sensitivity of Quadra was calculated for the $^{68}$Ga, $^{44}$Sc, $^{22}$Na and $^{124}$I radioisotopes. Finally, the sensitivity for the scandium isotope was compared to the sensitivities obtainable with the Total Body J-PET scanner, as well as with the modular J-PET prototype.

\parttitle{Results} 
The PLI sensitivities for the Quadra with the background radiation are estimated to 9.22(3), 10.46(4), 5.91(3), and 15.39(4) cps/kBq for the $^{44}$Sc, $^{68}$Ga, $^{22}$Na and $^{124}$I radioisotopes, respectively.
The highest sensitivity was obtained when the energy window for the deexcitation photon is adjacent to the energy window for the annihilation photons.
The determined PLI sensitivities with Quadra and the Total Body J-PET are in the order of sensitivities of standard PET imaging with the short axial field-of-view ($\sim$20 cm) PET scanners.

\parttitle{Conclusion}
The organ-wise PLI sensitivity of Quadra has been computed for the $^{68}$Ga, $^{44}$Sc, $^{22}$Na and $^{124}$I radioisotopes. A sensitivity gain by a factor of 150 was estimated relative to the modular J-PET system previously used for the first in-vivo PLI.

\end{abstract}


\begin{keyword}
\kwd{Positronium Lifetime Imaging}
\kwd{J-PET}
\kwd{Biograph Vision Quadra}
\kwd{Energy Window Technique}
\kwd{Sensitivity}
\end{keyword}


\end{abstractbox}
%

\end{frontmatter}

\newpage
\section{Background}
The standard procedure of Positron Emission Tomography (PET) provides a~metabolic image. Traditional Standardized Uptake Value (SUV) parameter quantifies the uptake of a given marker in tissues, enabling to identify regions of pathological metabolism \cite{Alavi2021,Schwenck2023,Bailey2005}. Nevertheless, current PET systems cannot yet take advantage of measuring the lifetime of the metastable positronium atom formed in the intra-molecular voids and the information that may be obtained from it \cite{Vandenberghe2020,Aide2022}. 
For~this purpose, a new PET-based imaging method has been conceived \cite{Moskal2020_IEEE} -- Positronium Lifetime Imaging (PLI) -- to provide complementary information about the imaged tissue.
In particular, mean lifetime or formation probability \cite{Moskal2025,Moskal2024,Moskal2022,Kubicz2023} constitute additional and qualitatively different contrast mechanisms from SUV \cite{Moskal2019}.
This novel type of imaging is recently gaining interest in both laboratory and hospital  research \cite{Moskal2024,Will2024} with its many challenges and new opportunities, e.g. analysis of positronium lifetime spectra \cite{Dulski2020,Shibuya2022}, PLI algorithms \cite{Qi2022,Shopa2023,Zhuo2024,Huang2024,Huang2024_2,Huang2025}, double tracer imaging \cite{Pratt2023,Fukuchi2021,Uenomachi2022,Beyene2023}. Although the requirement of three photon coincidence reduces sensitivity significantly, it may be resolved by another current trend towards the construction of long axial field-of-view PET scanners \cite{Mingels2024,Badawi2019,Alberts2023,Prenosil2022,Dai2023,Karp2020,Moskal2020}.

The first applications of both ex-vivo and in-vivo PLI were demonstrated by the Jagiellonian-PET (J-PET) collaboration \cite{Moskal2020_IEEE,Moskal2024,Dulski2021}. However, the first-ever clinical PLI image of the brain presented by the J-PET collaboration suffers from the low statistics due to the utilized radionuclide ($^{68}$Ga) and low scanner’s sensitivity \cite{Moskal2024}. Firstly, the pharmaceutical labeled with $^{68}$Ga radionuclide has a low branching ratio (1.2\% \cite{Ga_decay}) for $\beta^{+}$ decay followed by a deexcitation through the emission of a prompt gamma \cite{Das2023}. The administration of dedicated radiopharmaceuticals, such as $^{44}$Sc radionuclide, can enhance statistics by about 76 times \cite{Moskal2020,Sitarz2020,Matulewicz2021}. Secondly, the patient was imaged with the prototype of the J-PET tomograph - single ring only modular J-PET with 50 cm axial field-of-view (AFOV) constructed with organic scintillators \cite{Moskal2024,Faranak2024,Faranak2023}. Its sensitivity for metabolic and positronium imaging amounts to 3.4 and 0.06 cps/kBq \cite{Moskal2024}.
Therefore, in this study we investigate a crystal-based system with higher interaction probability of photons.

Most of the clinically certified PET tomographs are acquiring data in the form of sinograms or list-mode pairs, which limits raw data to only double coincidences. However, in order to perform positronium imaging, a specific type of triple coincidence is required. The type of acquisition, which allows for the registration of all interactions is referred to as triggerless mode \cite{Korcyl2018,Moskal2014} or singles mode \cite{Will2024}. Apart from the J-PET system, which works in the triggerless data acquisition mode, the PennPET Explorer in Philadelphia, USA \cite{Dai2023,Karp2020} as well as the recently upgraded Biograph Vision Quadra (hereinafter referred to as Quadra) tomograph from Siemens Healthineers \cite{Will2024} have the possibility of singles mode acquisition.

\begin{figure}[h!]
    \centering
    \includegraphics[width=0.7\textwidth]{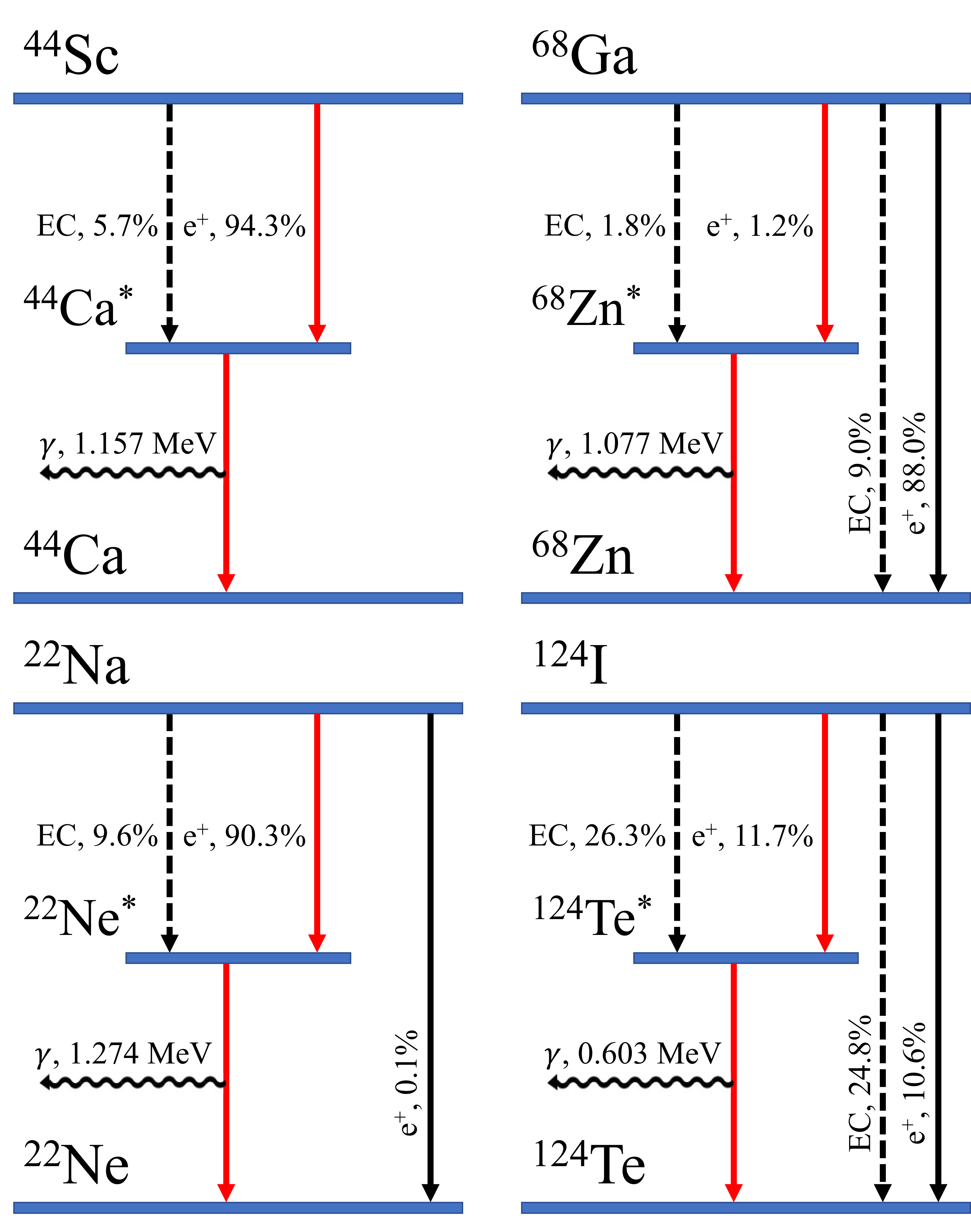}
    \caption{Schematic of selected decay modes relevant to PET imaging for the four investigated radioisotopes: $^{44}$Sc, $^{68}$Ga, $^{22}$Na and $^{124}$I. Positron emission is denoted by e$^{+}$, whereas EC indicates the contribution from electron capture \cite{Ga_decay,Sc_decay,Na_decay,I_decay}.}
    \label{fig:isotopes}
\end{figure}

In this simulation-based study, a feasibility of positronium lifetime imaging with Quadra was investigated and compared to the J-PET scanners. We focus on PLI sensitivity, which has been identified as a major issue in the images previously presented by the J-PET collaboration and is expected to significantly improve with the crystal-based and Total-Body tomographs. Four isotopes were explored in this work: $^{68}$Ga, $^{44}$Sc, $^{22}$Na and $^{124}$I \cite{Moskal2020,Dulski2021,Sitarz2020,Matulewicz2021,Lang2014,Thirolf2015}. The decay scheme of these radionuclides are shown in Figure \ref{fig:isotopes}. The isotope of gallium was chosen due to its common use in PET imaging. The scandium radioisotope is considered due to its feasibility for positronium imaging \cite{Das2023}. Despite the present lack of clinical imaging applications, the $^{22}$Na isotope was chosen due to its exploitation in basic research \cite{Bass2023}. Finally, the iodine isotope was chosen for its diagnosis and treatment capabilities of thyroid cancer \cite{Phan2008}.

\section{Methods}

\subsection{Positronium imaging}
The pharmaceuticals administered to the patients during PET scan are labeled with specific radioisotopes, which undergo $\beta^{+}$ decays. After thermalization, the emitted positron may directly annihilate with an electron from the surrounding material or create a bound, metastable atom consisting of e$^{+}$ and e$^{-}$ called positronium (Ps) \cite{Bass2023}, which can be formed in one of the two states: para-positronium (p-Ps) or ortho-positronium (o-Ps), annihilating into an even or odd numbers of photons, respectively. The difference in lifetime of p-Ps and o-Ps states is especially profound in vacuum, where their respective lifetimes are 125 ps and 142 ns. Particularly, in medical purposes the ortho-positronium lifetime plays a significant role, due to its susceptibility to environmental conditions \cite{Moskal2019}. Processes such as o-Ps to \mbox{p-Ps} conversion or pick-off effectively shorten the Ps lifetime \cite{Bass2023}. These alterations provide information about the intramolecular structure \cite{Moskal2022,Kubicz2023,Moskal2019,Dulski2021,Bass2023,Jasinska2017,Moskal2021,Bass2019}.

\begin{figure}[h!]
    \centering
    \includegraphics[width=0.95\textwidth]{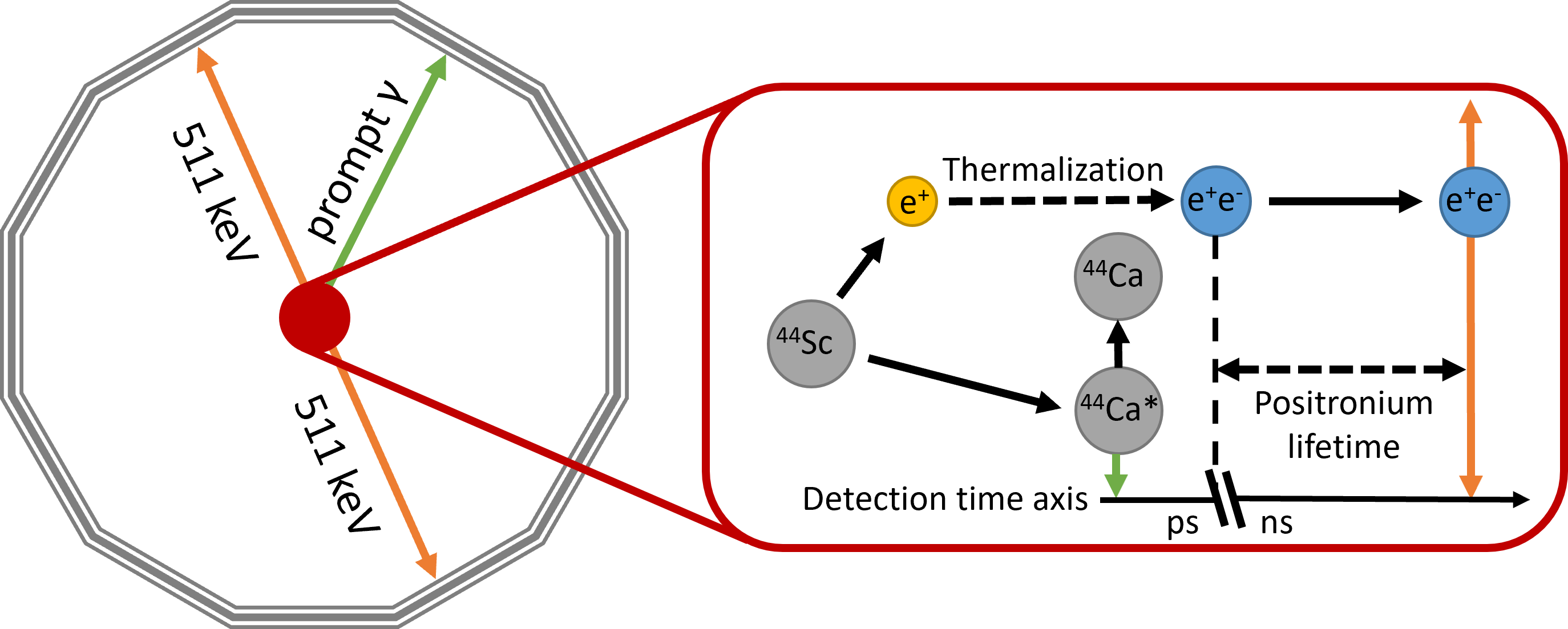}
    \caption{Scheme of the principle of operation of positronium lifetime imaging based on the $^{44}$Sc radioisotope. $^{44}$Sc undergoes a $\beta^{+}$ decay with the emission of an electron neutrino and a positron. The radionuclide transforms into the excited state $^{44}$Ca*, which then deexcites with the emission of a 1157 keV prompt gamma (marked on the main figure and the zoom with the green arrow) \cite{Moskal2020}. The emitted positron, after thermalization, may form together with an electron a bound state called positronium. In case an ortho-positronium (o-Ps) state has been created, a conversion or pick-off process results in annihilation and back-to-back emission of two 511 keV photons (marked on the main figure and the zoom with the orange arrows). Next, the annihilation photons and prompt   gamma can be registered by the scintillator detectors placed in the PET scanner (gray frame). Since the expected lifetime is in order of ns, while the deexcitation time of $^{44}$Ca* and the thermalization process in ps, the prompt gamma indicates the time of positronium creation \cite{Bass2023,Kisielewska2020,Kisielewska2019}.}
    \label{fig:pop}
\end{figure}

PLI has been invented in order to take advantage of positronium lifetimes alteration as a diagnostic biomarker \cite{Moskal2020_IEEE,Moskal2024,Moskal2022,Kubicz2023}. It employs use of a specific radioisotopes, which deexcite with the emission of a prompt gamma after the $\beta^{+}$ decay. A pictorial representation of PLI's principle of operation is presented in Figure \ref{fig:pop}. Since both deexcitation and thermalization phenomena are in a similar time range (in the order of ps) and expected lifetimes are in the order of ns, registration of a prompt gamma can be treated as a timestamp of Ps formation \cite{Kisielewska2020,Kisielewska2019}. Hence, positronium lifetime may be estimated as a difference between prompt gamma emission and positronium annihilation time.

\subsection{Simulation software}
Simulation of radioactive $\beta^{+}$ decays and deexcitation of the daughter nuclei, together with tracing of their products interacting within PET tomographs, was performed with the GATE (Geant4 Application for Tomographic Emission) software (v.9.0) \cite{Sarrut2021,Sarrut2022}. It is an established open source software developed by the OpenGATE Collaboration for Monte-Carlo studies in medical imaging. GATE enables the generation of three positronium decay schemes: p-Ps decaying into two photons, o-Ps decaying into three photons and a weighted mix of both p-Ps and o-Ps. However, it does not simulate interactions of atoms with medium such as conversion or pick-off process that shorten the o-Ps lifetime significantly and result in annihilations into two photons. Hence, in this study we simulated p-Ps decay into two photons and emission of prompt gamma with uncorrelated direction with respect to the propagation of annihilation photons, assuming the default lifetime in vacuum.

\subsection{Studied PET scanners}
This research focuses mainly on the Biograph Vision Quadra (Quadra) tomograph from the Siemens Healthineers \cite{Prenosil2022}, with a comparison to results achieved and possible to achieve with the modular J-PET prototype \cite{Moskal2024,Dulski2021,Faranak2024,Faranak2023,Parzych2023} and the Total Body J-PET scanner~\cite{Kowalski2021,Dadgar2023}. Counterparts of all of the scanners were simulated with the GATE software.

\subsubsection{Biograph Vision Quadra}
The Quadra tomograph \cite{Prenosil2022} utilizes 20$\times$3.2$\times$3.2 mm$^{3}$ Lutetium Oxyorthosilicate (LSO) scintillator crystals. An array of 5×5 crystals forms a mini block, where 8 of them (2 in axial and 4 in transaxial direction) constitutes a detector block. Further, 16 blocks (2 in transaxial and 8 in axial direction) form one detector electronics assembly (DEA) (see Figure \ref{fig:geometry}). 19 DEA create one ring with 82 cm inner diameter. In total, Quadra consists of 4 rings providing a 106 cm AFOV (see Table \ref{tab:geom}).

During simulation all deposited energies were smeared with a normal distribution. Following the inverse square law, its full width at half maximum (represented by resolution) was recalculated assuming a 9\% resolution at an energy equal to 511~keV~\cite{Prenosil2022}.

\subsubsection{Modular J-PET}
The modular J-PET scanner \cite{Faranak2024,Faranak2023} is built from 24$\times$6$\times$500 mm$^{3}$ BC-404 plastic scintillator strips. Each strip is read on both ends with electronic systems. 13 axially arranged strips form a module (see Figure \ref{fig:geometry}), and 24 modules constitute a one layer ring scanner with an inner diameter of 74 cm. The modular J-PET has AFOV of 50~cm (see Table \ref{tab:geom}).
It is equipped with the trigerless data acquisition \cite{Korcyl2018,Korcyl2014} enabling multi-photon tomography \cite{Dulski2021,Gajos2021} and applications of the degree of annihilation photons polarization \cite{Czerwinski2024,Kumar2025}. Its energy resolution was set to 23\% at energy equal to 200 keV \cite{Moskal2014}.

\subsubsection{Total Body J-PET}
\begin{figure}[h!]
    \centering
    \includegraphics[width=0.95\textwidth]{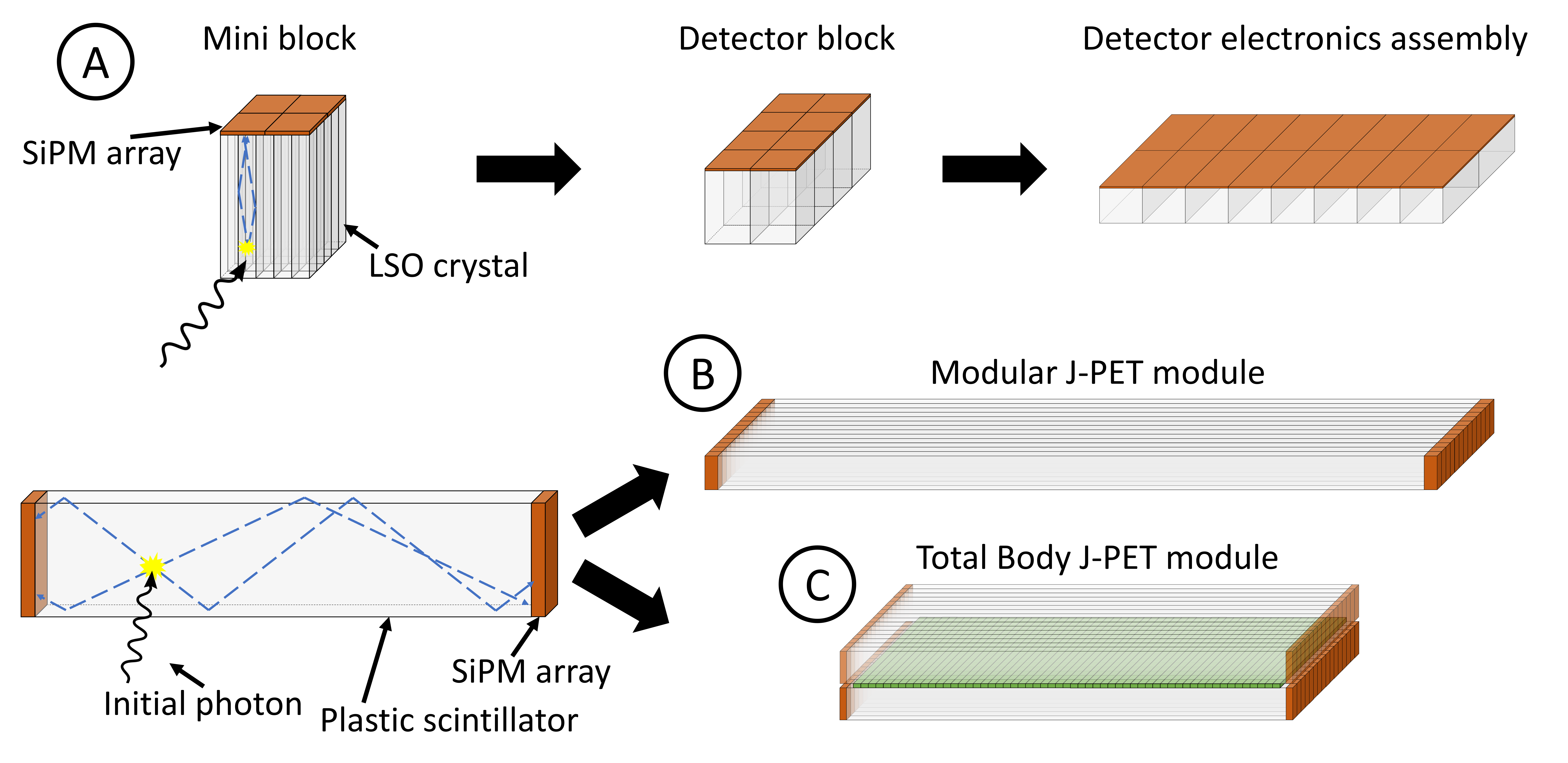}
    \caption{(A) Scheme of the mini block, detector block and detector electronics assembly, which constitute building parts of the Biograph Vision Quadra.  
    (B)~Modular and (C) Total Body J-PET detection modules with plastic scintillators (green color marks the wavelength shifters layer).}
    \label{fig:geometry}
\end{figure}

The simulated Total Body J-PET consists of 7 rings, each built from two layers of cylindrically arranged BC-408 plastic scintillators strips with dimensions of 30$\times$6$\times$330 mm$^{3}$ \cite{Kaplon2023}. Every ring is constructed from 24 modules, each with 16 scintillators per layer, per ring (see Figure \ref{fig:geometry}). In total, the tomograph spans a 243~cm AFOV, with an inner diameter of 83 cm. The whole system will be set at the cross-staged gantry  \cite{Kaplanoglu2023}. Moreover, an additional perpendicular layer of BC-482A wavelength shifters \cite{Smyrski2017,Smyrski2014,Georgadze2023} is situated between the two layers of plastic strips within each module (see Table \ref{tab:geom}). The purpose of this layer is to further enhance the axial resolution of reconstructed interaction position of photons \cite{Kowalski2021,Smyrski2017}. The~energy resolution for this systems was set to 23\% at an energy equal to 200 keV \cite{Moskal2014}.

\begin{table}[h!]
\caption{Geometries of the PET scanners utilized in this work. The \textit{Scintillator per module} row refers to the amount of scintillators present in the detector electronics assembly in Biograph Vision Quadra and module in J-PET tomographs.}
 \normalsize
\begin{tabular}{|c|c|c|c|}
\hline
Scanner &
  \begin{tabular}[c]{@{}c@{}}Biograph Vision\\ Quadra\end{tabular} &
  Modular J-PET &
  \begin{tabular}[c]{@{}c@{}}Total Body\\ J-PET\end{tabular} \\ \hline
\begin{tabular}[c]{@{}c@{}}Scintillator\\ material\end{tabular}        & LSO & BC-404 & BC-408 \\ \hline
\begin{tabular}[c]{@{}c@{}}Scintillator\\ dimensions {[}mm$^{3}${]}\end{tabular} &
  20$\times$3.2$\times$3.2 &
  24$\times$6$\times$500 &
  30$\times$6$\times$330 \\ \hline
\begin{tabular}[c]{@{}c@{}}Scintillator\\ per module*\end{tabular} &
  3200 &
  13 &
  \begin{tabular}[c]{@{}c@{}}32\\ (+ WLS layer)\end{tabular} \\ \hline
\begin{tabular}[c]{@{}c@{}}Modules\\ per ring\end{tabular}             & 19  & 24      & 24      \\ \hline
\begin{tabular}[c]{@{}c@{}}Rings per\\ scanner\end{tabular}            & 4   & 1       & 7       \\ \hline
\begin{tabular}[c]{@{}c@{}}Inner\\ diameter {[}cm{]}\end{tabular}      & 82  & 74      & 83      \\ \hline
\begin{tabular}[c]{@{}c@{}}Axial\\ field-of-view {[}cm{]}\end{tabular} & 106 & 50      & 243     \\ \hline
\end{tabular}
 \label{tab:geom}
\end{table}

\subsection{Energy window technique}
\begin{figure}[h!]
    \centering
    \includegraphics[width=0.95\textwidth]{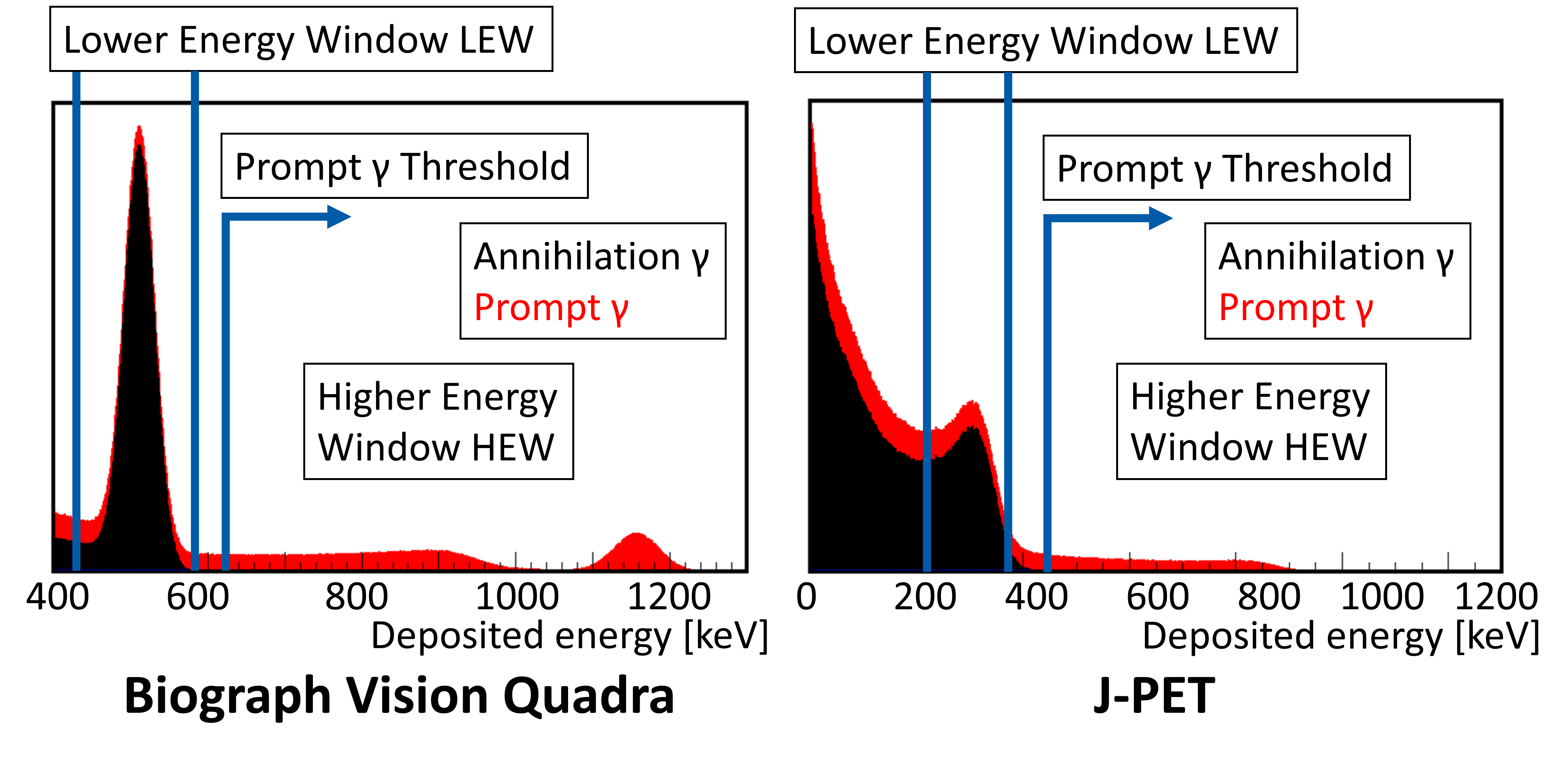}
    \caption{Exemplary simulated energy deposition spectra for the Biograph Vision Quadra (left) and the modular J-PET (right). Red part of the spectra correspond to prompt gammas and black to annihilation photons. Moreover, exemplary positioning of Lower Energy Window for annihilation $\gamma$ and Higher Energy Window for deexctitation $\gamma$ (with its bottom threshold marked as Prompt $\gamma$ Threshold) are denoted and marked with blue lines. The energy resolution of 9\% (at 511 keV) was set for the Biograph Vision Quadra \cite{Prenosil2022} and 23\% (at 200 keV) for the modular J-PET \cite{Moskal2014}.}
    \label{fig:ew}
\end{figure}

The conventional metabolic imaging in crystal-based PET scanners focuses on registration of the 511 keV photons from e$^{+}$e$^{-}$ annihilation, which interact within the scintillators via the photoelectric effect. To that end, registration of interactions is filtered via the deposited energy, which is centered around 511 keV photopeak \cite{Conti2016}. The utilization of such an energy window (EW) allows also for immediate rejection of photons which underwent any scattering (e.g. in the patient's body), and whose involvement degrades the quality of the imaging. However, in PLI, it is necessary to register also a prompt gamma originating from radionuclide, whose initial energy for the most isotopes is higher than 511 keV \cite{Das2023}. Likewise, their photopeaks are situated partially or fully outside of the EW. Therefore, a second energy window needs to be introduced to register and distinguish prompt gammas. Since the EW for annihilation photons covers their highest possible energy depositions, in principle, EW for deexcitation photons may start right after the previous one and its upper limit may be left open. Hence, utilization of only the energy threshold (ET) for prompt gammas may be enough. We assume no external sources of radiation such as internal radiation (e.g. from $^{176}$Lu in LSO crystals \cite{Teimoorisichani2022}) or cosmic rays. In~any case, utilization of EW and ET lead to inclusion of scattered prompt gammas, worsening the quality of gathered data. A possible approach to find the proper placement of the energy window for prompt gamma was shown in reference \cite{Parzych2023}. we adapt their nomenclature (see Figure \ref{fig:ew}):
\begin{itemize}
    \item energy window for annihilation photons will be denoted as the Lower Energy Window (LEW);
    \item energy window for prompt gamma will be called the Higher Energy Window (HEW);
    \item bottom energy threshold of the HEW will be named the Prompt $\gamma$ Threshold.
\end{itemize}
Quadra has a well established LEW from 435 keV to 585 keV \cite{Prenosil2022}. For isotopes with well distinguished annihilation and deexcitation photopeaks, the investigation is necessary only for the Prompt $\gamma$ Threshold. However, in the case of $^{124}$I isotope, which photopeaks partially overlap due to prompt gamma energy equal to 603 keV, the modification to LEW was also inspected. There, the upper limit of LEW was lowered, allowing the Prompt $\gamma$ Threshold to cover more of deexcitation photopeak.

In the case of the J-PET tomographs, which are constructed from the plastic scintillators, the principle behind PET detection is switched to the Compton-based interactions \cite{Moskal2014}. Nevertheless, the same reasoning as abovementioned can be deduced by exchanging photopeak region to the Compton edge region. As shown in \cite{Parzych2023}, the optimal Lower Energy Window for the $^{44}$Sc radioisotope spans the 200 - 370 keV energy region with a closely following Prompt $\gamma$ Threshold. The lower limit on LEW was previously studied \cite{Moskal2014} and doesn’t depend on the utilized radioisotope (annihilation photons energy are independent of the source).

\subsection{Event selection}
GATE v9.0 does not support creation of triple (2 annihilation + 1 deexcitation) coincidences. Consequently, only the information on the level of Singles was saved and further analyzed in post-processing. The analysis takes as an input series of Singles sorted in time. Further, it imposes a time window to find triple coincidences. The first time window is opened with first detection. All of the next windows are opened with the subsequent detections, which did not fall into the already opened time window. If, during the time window, exactly two interactions fell into the LEW and one into the HEW, a candidate for the triple coincidence was exported.

This study focuses on the determination of system sensitivity for PLI which is independent of the positronium lifetime. Therefore, for simplicity, a default positronium lifetime in vacuum was simulated and the time windows analogical to the case of metabolic imaging were chosen: 4.7~ns and 3~ns (dictated by the geometrical acceptance) for Quadra and J-PET, respectively \cite{Prenosil2022,Parzych2023}. 
During metabolic imaging both registered photons are emitted at the same time. However, in PLI the time window should be dictated by the lifetime of positronium in the investigated object. The mean positronium lifetime for different tissues like cardiac myxoma, adipose tissue, glioblastoma tumor, salivary glands and healthy brain tissues ranges from about 1.4~ns to 4~ns \cite{Moskal2024,Kubicz2023,Dulski2021,Jasinska2017,Ahn2021,Jean2006,Chen2012,Avachat2024,Jasinska2017p2,Zgardzinska2020,Moyo2022,Karimi2023}. 
Therefore, in experimental positronium lifetime imaging the time window has to be extended, e.g. to 10 ns.

\subsection{Data analysis}
Each registered candidate for the triple coincidence may be categorized based on the discrimination of the correct photon and on annihilation-deexcitation event history. Firstly, overlapping energy spectra of annihilation and prompt gamma may cause a misidentification of annihilation photon in LEW or prompt gamma in HEW, creating improper candidate triples.
The number of correct triple candidates $N_{triples}^{correct}$ (identified with the purity $P_{triple}$) describes a situation when 2 detections in LEW are indeed coming from annihilation and a single detection in HEW from deexcitation, following definition:
\begin{equation}
    P_{triple} = \frac{N_{triples}^{correct}}{N_{triples}} ,
\label{eq:triple}
\end{equation}
where $N_{triples}$ denotes number of all registered triple coincidences.
Secondly, based on annihilation-deexcitation event history, correct triple candidates may be sorted into standard coincidence types: true (with purity $P_{t}$), scattered (with purity $P_{s}$) and accidental (with purity $P_{a}$). The scattered coincidence arises when at least one of the detected photons underwent other interaction in the imaged object or detector prior to current registration. Accidental coincidence denotes a situation when at least one gamma is originating from a different annihilation-deexcitation event. The purity of true coincidences is defined as:
\begin{equation}
    P_{t} = \frac{N_{triples}^{correct, \ true}}{N_{triples}^{correct}} ,
\label{eq:triple}
\end{equation}
where $N_{triples}^{correct, true}$ corresponds to all correct triple candidates that are not falling into scattered or accidental coincidence criterion.

For the optimization of positioning of the HEW the following figure of merit (FoM) was considered:
\begin{equation}
    FoM = P_{triple} \times P_{t} \times \varepsilon^{rel}_{pr} ,
\label{eq:fom}
\end{equation}
where $\varepsilon^{rel}_{pr}$ corresponds to the relative efficiency of prompt gamma registration (defined in equation \ref{eq:eff}).

After determining the most optimal energy windows, an organ-wise sensitivity for PLI was estimated. The organ-wise sensitivity is defined as the mean of sensitivities of slices within the $\pm$4.5 cm region around the scanner’s center. 
The simulation was performed using a 183 cm linear source located at the central axis of the tomograph with 1 MBq total activity. 
The sensitivity of the $i^{th}$ slice $\varepsilon_{i}$ was calculated based on the NEMA standards \cite{NEMA} as a ratio of registered true triple coincidences falling into true coincidence category originating from within the slice ($N_{triples, \ i}^{correct, \ true}$) to the activity present within it according to the following formula:
\begin{equation}
    \varepsilon_{i} = \frac{R_{i} \times L}{d \times A} ,
\label{eq:sensitivity}
\end{equation}
where $L$ is the source length, $d$ is the width of the slice, and $A$ is the activity. The rate $R_{i}$ of registration in counts per second is defined as $N_{triples, \ i}^{correct, \ true}$ divided  by the duration of the measurement. Simulation took also into account accidental coincidences, as well as possible scatterings in the detector.

\section{Results}

\subsection{Efficiency of the energy window method}
\begin{figure}[h!]
    \centering
    \includegraphics[width=0.5\textwidth]{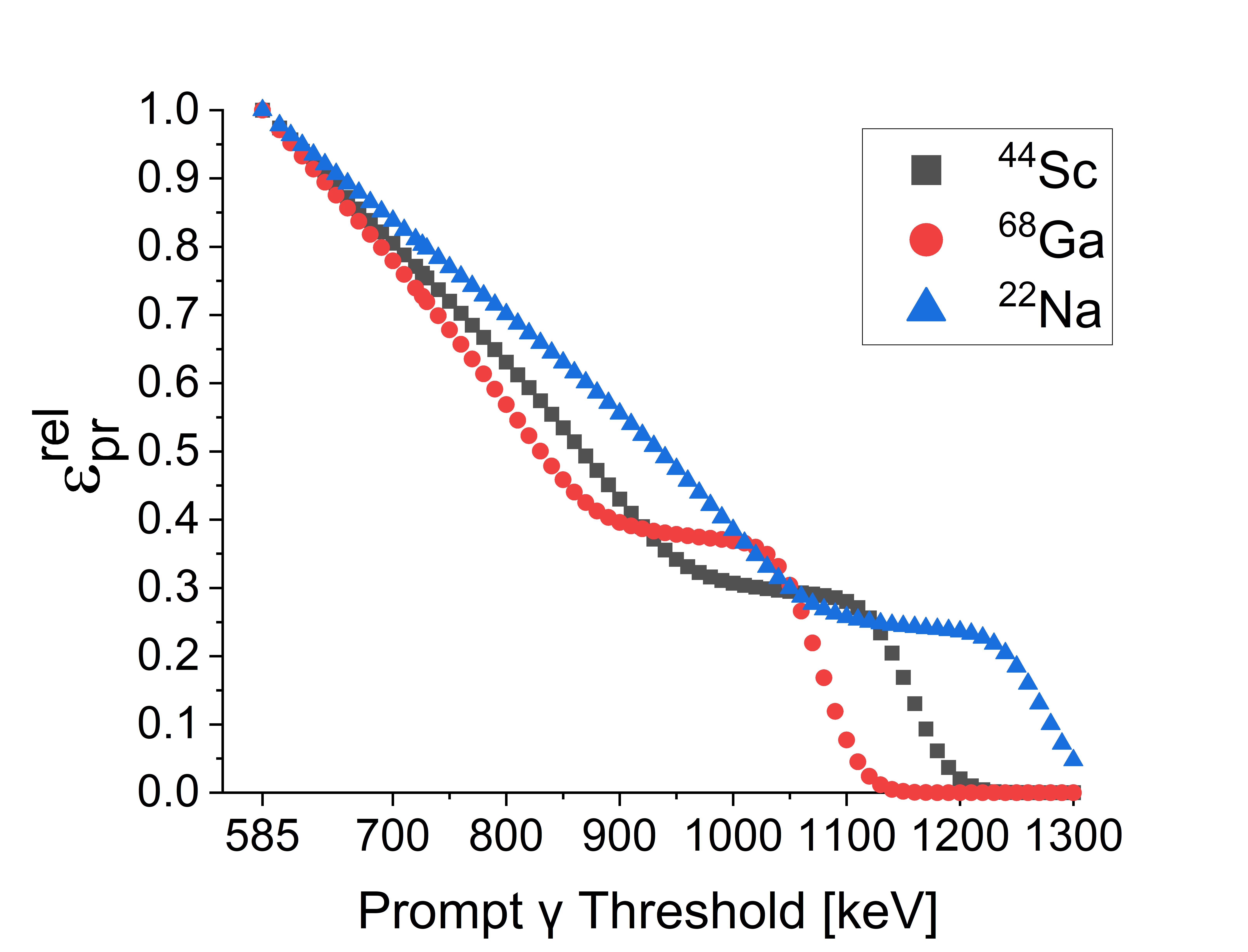}
    \caption{Dependence of prompt gamma detection relative efficiency ($\varepsilon^{rel}_{pr}$) on the Prompt $\gamma$ Threshold placement in the Biograph Vision Quadra for $^{44}$Sc (square), $^{68}$Ga (circle) and $^{22}$Na (up-triangle) radioisotopes. 
    Total of 10$^{8}$ interactions of prompt photons in the detector were simulated.
    }
    \label{fig:eff}
\end{figure}

In case of the metabolic imaging, the standard energy window is used to register only annihilation photons. However, if the $\beta^{+}$ decay is followed by a deexcitation, as in the case in PLI, an incorrect photon might be registered within the respective EW. Since the energy deposition spectra of both types of gammas may overlap, the purity of EW selection is degraded. Since Quadra has already a well established LEW for annihilation photons (435-585 keV), we only investigate the positioning and impact of HEW. Figure \ref{fig:eff} shows the efficiency for choosing the prompt gamma in the HEW for the given energy thresholds. The relative efficiency is defined as:
\begin{equation}
    \varepsilon^{rel}_{pr} = \frac{N_{pr}}{N_{pr}^{585 keV}} ,
\label{eq:eff}
\end{equation}
where $N_{pr}$ corresponds to amount of registered prompt gammas as a function of Prompt $\gamma$ Threshold and $N_{pr}^{585 keV}$ denotes the reference at 585 keV. Such defined relative efficiency is equal to 1 for the Prompt $\gamma$ Threshold starting at 585 keV, right after LEW of Quadra. 
In all cases, the efficiency of prompt registration is falling rapidly, with a small plateau region for energies above the prompt gamma's Compton edge and below its photopeak. The $^{124}$I radioisotope will be investigated separately due to modifications in its LEW.

\subsection{Characteristics of triple coincidences}
\begin{figure}[h!]
    \centering
    \includegraphics[width=0.5\textwidth]{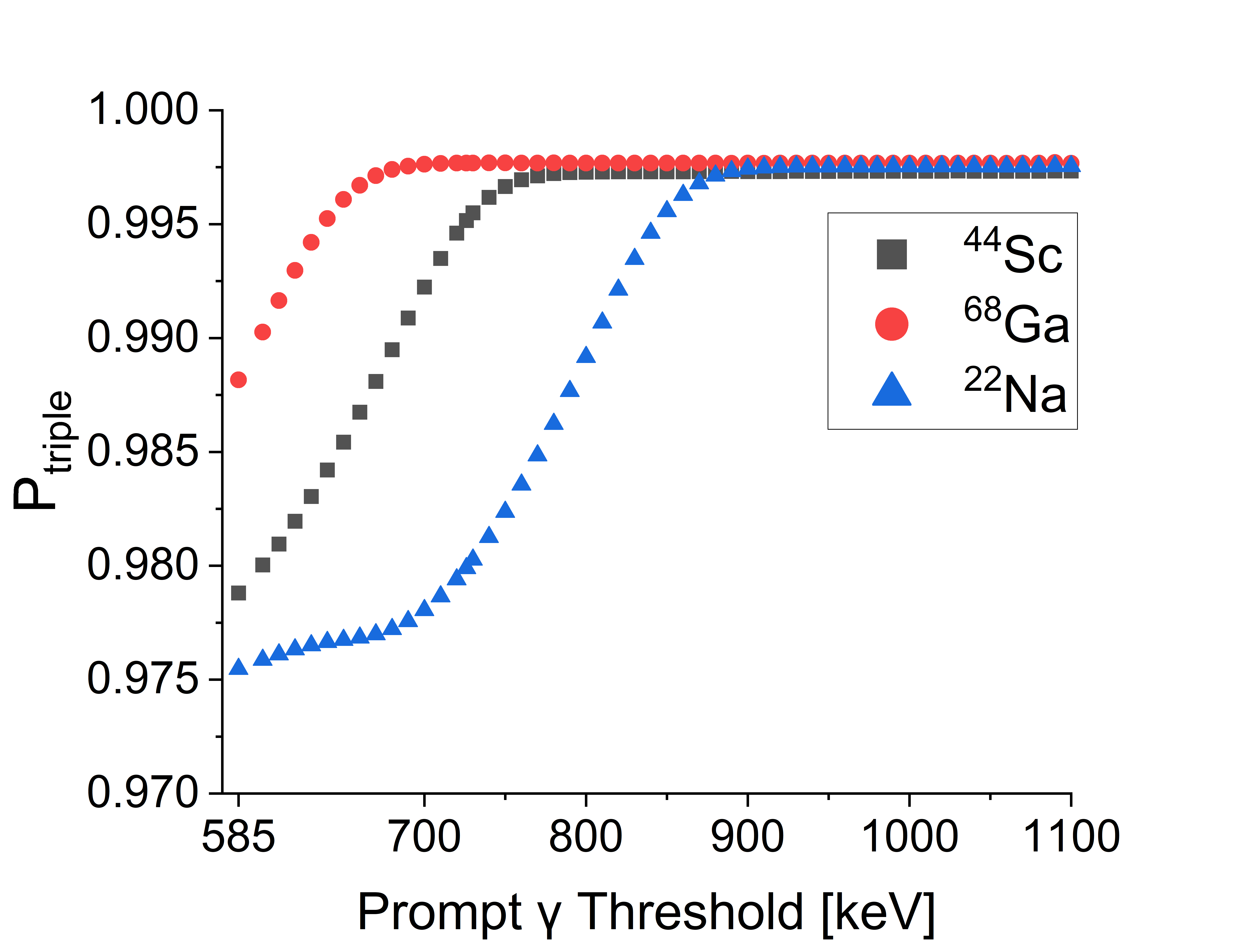}
    \caption{Dependence of correct candidate triple selection purity ($P_{triple}$) on the Prompt $\gamma$ Threshold for the Biograph Vision Quadra. 
    Results obtained for $^{44}$Sc radioisotope are shown with squares, for $^{68}$Ga with circles and for $^{22}$Na with up-triangles.}
    \label{fig:triple}
\end{figure}

\begin{figure}[h!]
    \centering
    \includegraphics[width=0.5\textwidth]{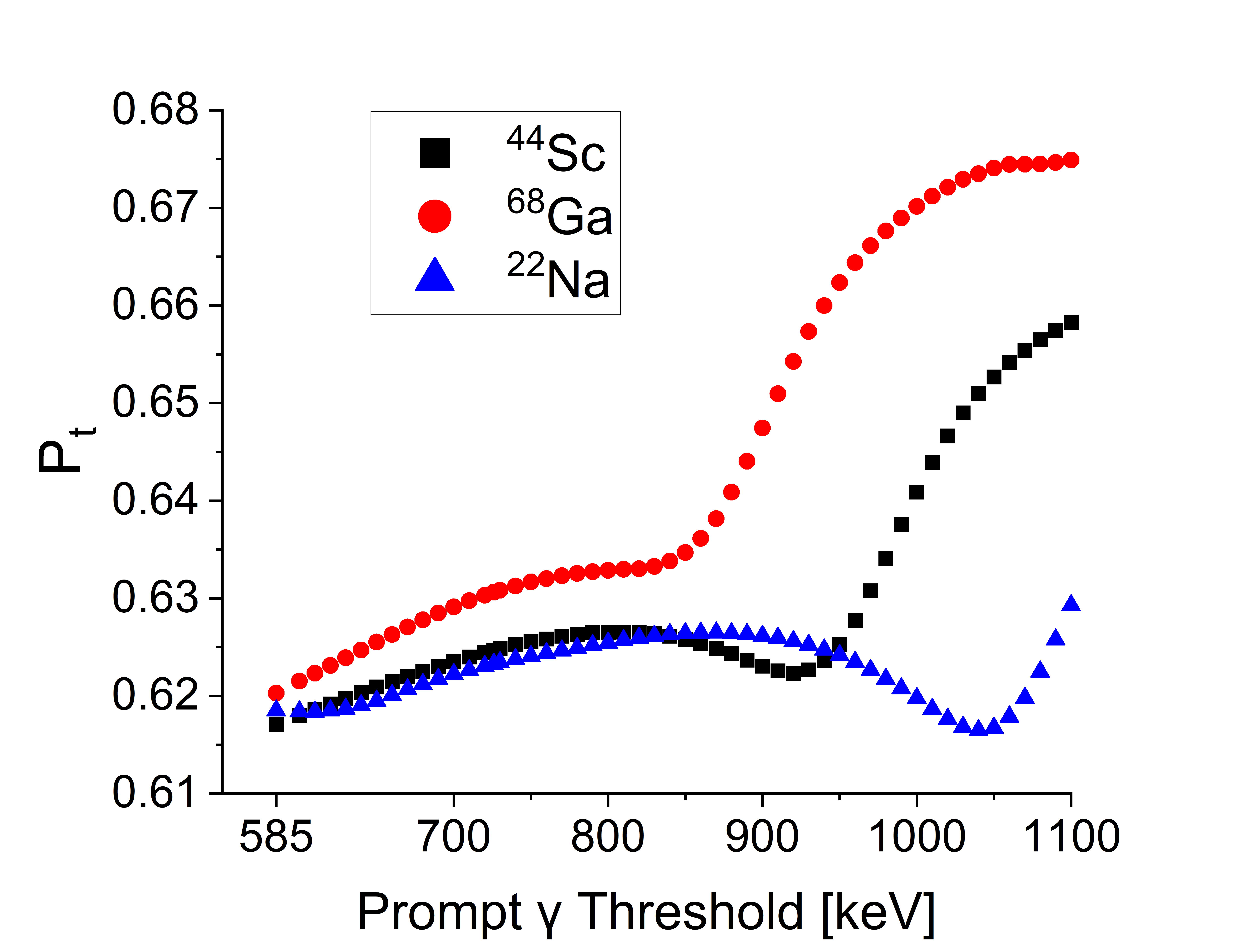}
    \caption{Dependence of true coincidence registration probability from within correct candidate triples ($P_{t}$) on the Prompt $\gamma$ Threshold for the Biograph Vision Quadra. The true coincidence requires detection of unscattered photons originating from the same annihilation-deexcitation event. Results obtained for $^{44}$Sc radioisotope are shown with squares, for $^{68}$Ga with circles and for $^{22}$Na with up-triangles.}
    \label{fig:true}
\end{figure}

\begin{figure}[h!]
    \centering
    \includegraphics[width=0.8\textwidth]{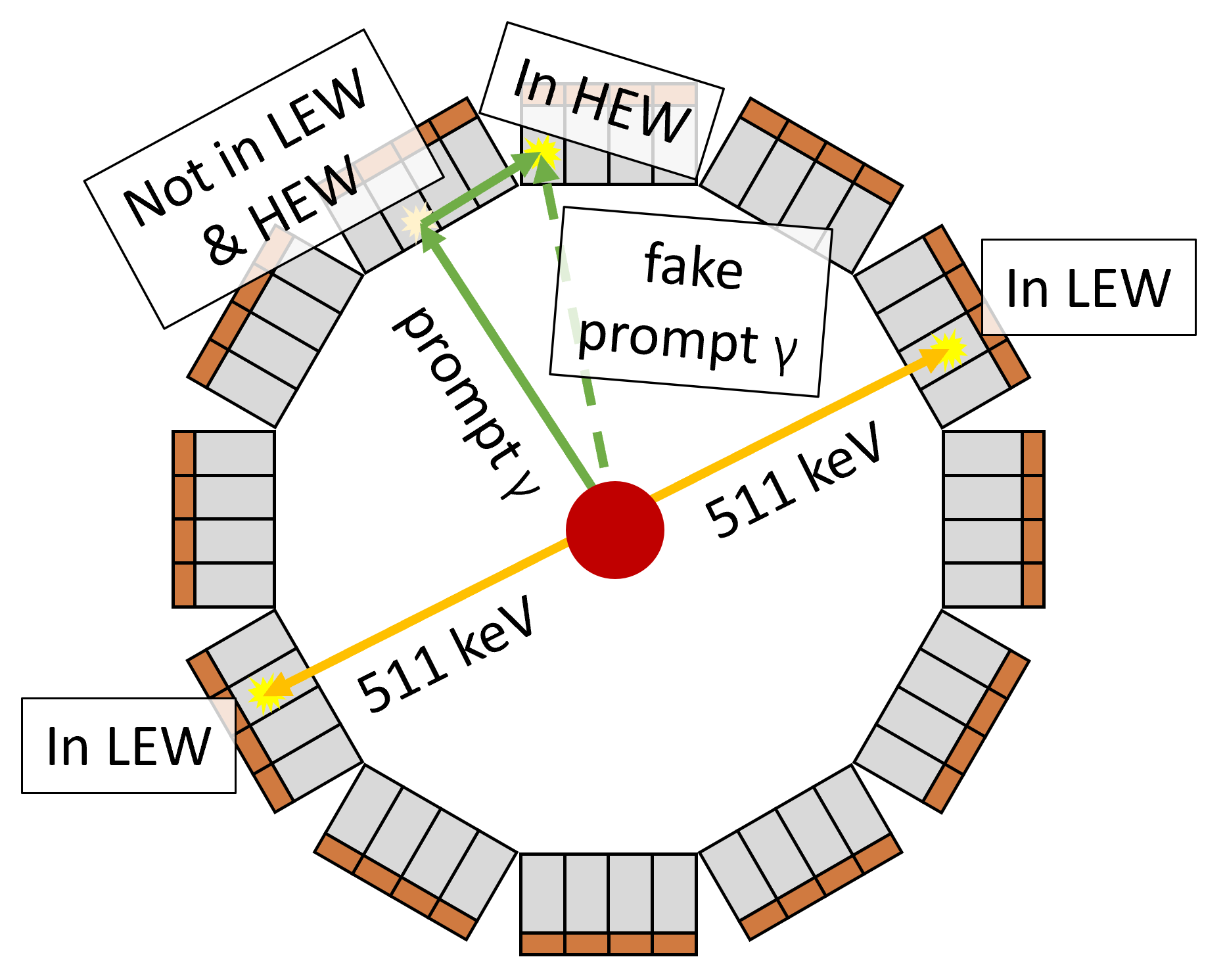}
    \caption{Exemplary signature of correct triple coincidence falling into scattered coincidence category due to reconstruction of a fake prompt gamma.}
    \label{fig:scatter}
\end{figure}

Figure \ref{fig:triple} shows the purity of selecting the true candidate triples ($P_{triple}$).
Results show high ($>$97\%) purity of selecting such a true candidate with quick saturation. As shown in Figure \ref{fig:true}, the purity of registering true coincidences within the correct candidate triples slightly increases (about 1 percent) in the first half of Prompt~$\gamma$~Threshold range. The following drop of $P_{t}$ signifies the higher impact of small-angle scattered photon registrations. The final rise of $P_{t}$ corresponds to the reduction in detection of Compton interactions and scattered photons (exemplary signature shown in Figure \ref{fig:scatter}) in favor of photoelectric effect based registration of original prompt gamma. The impact of the accidental coincidences remains constant in the whole range and amounts to less than 2\%, which is cause by the low simulated activity (1 MBq).

\begin{figure}[h!]
    \centering
    \includegraphics[width=0.5\textwidth]{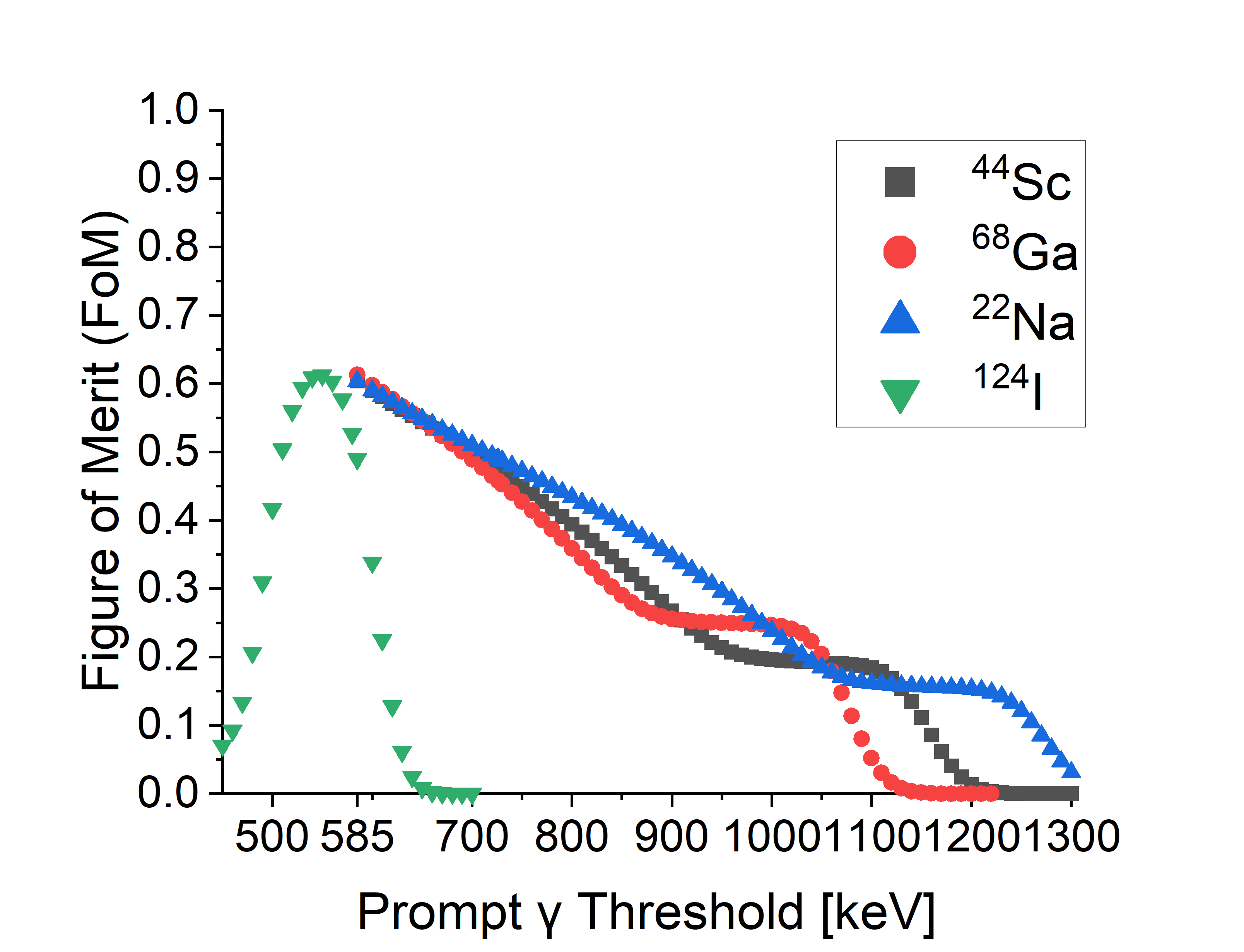}
    \caption{Dependence of introduced figure of merit (defined in eq. \ref{eq:fom}) on the Prompt $\gamma$ Threshold for the Biograph Vision Quadra. Results for $^{44}$Sc (square), $^{68}$Ga (circle) and $^{22}$Na (up-triangle) radioisotopes are shown.
    In case of $^{124}$I radioisotope (down-triangle), its definition is extended to take into account the purity of prompt gamma choice (eq. \ref{eq:fom_I}).}
    \label{fig:fom}
\end{figure}

To define the optimization of the HEW the FoM from equation \ref{eq:fom} was determined (Fig. \ref{fig:fom}). This variable combines the purity of choosing correct candidate triples (Fig. \ref{fig:triple}), purity of registering the true coincidence (Fig. \ref{fig:true}), and relative efficiency for qualifying prompt gamma (Fig. \ref{fig:eff}). As all of the parameters are in range of 〈0,1〉, the figure of merit takes values in the 〈0,1〉 region as well. Higher values indicate better quality of data, not contaminated with scattered gammas. The resulting shape is mostly dictated by the relative efficiency ($\varepsilon^{rel}_{pr}$), with constant downward trend, suggesting best performance for HEW adjacent to LEW.

\subsection{Case of $^{124}$I radioisotope}
\begin{figure}[h!]
    \centering
    \includegraphics[width=0.7\textwidth]{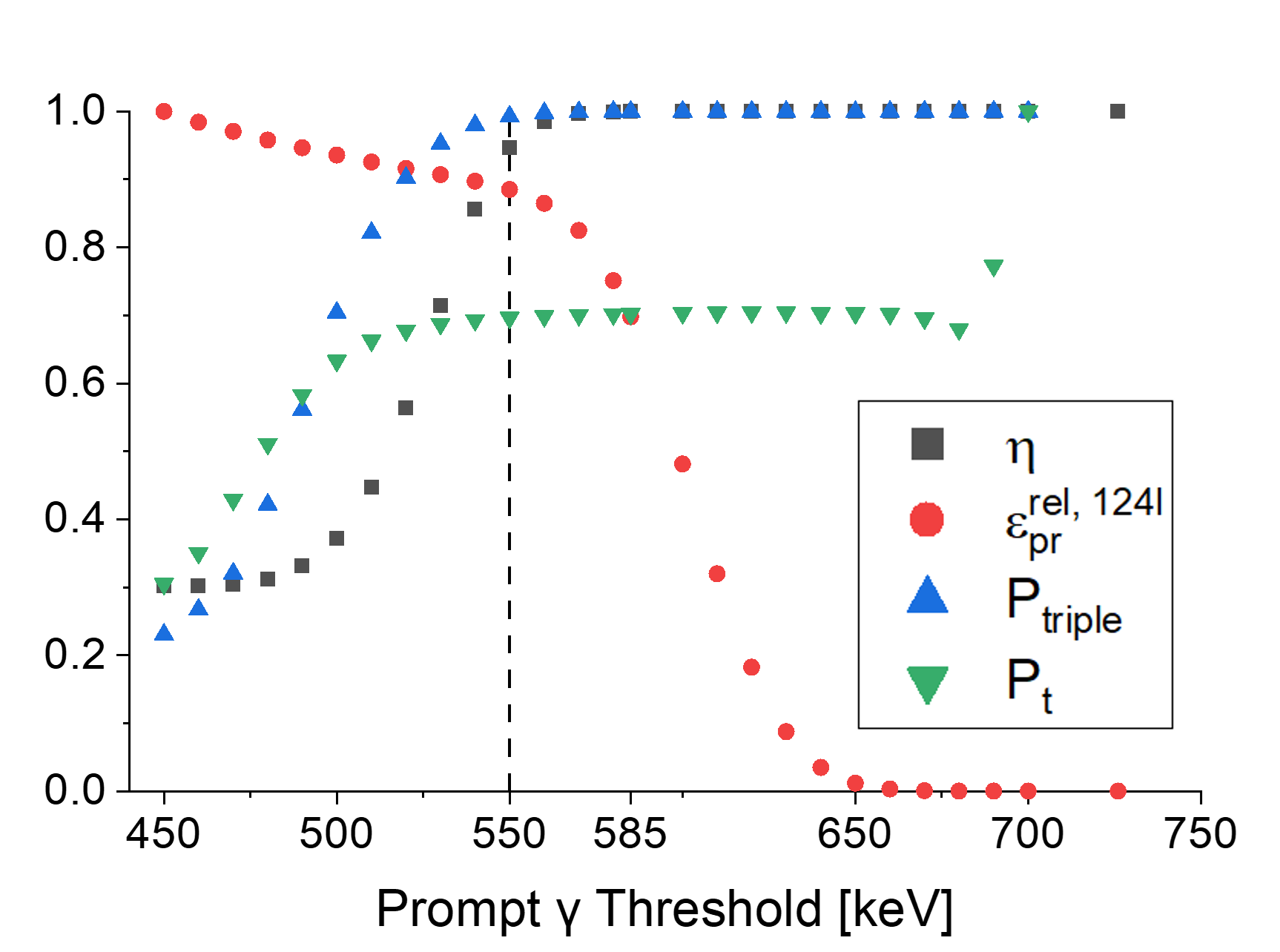}
    \caption{Dependence of purity ($\eta$, square) and relative efficiency ($\varepsilon^{rel}_{pr}$, circle) of prompt gamma detection in HEW, purity of true candidate triple ($P_{triple}$, up-triangle) and true coincidence selection ($P_{t}$, down-triangle) on the Prompt $\gamma$ Threshold for $^{124}$I radioisotope examination in the Biograph Vision Quadra. The resulting figure of merit is shown in Figure \ref{fig:fom}}
    \label{fig:124I}
\end{figure}

For the iodine isotope ($^{124}$I), the photopeaks of annihilation and deexcitation photon overlap at each other significantly. Therefore, the upper limit of LEW was set free for the investigation of the optimal HEW, which might have started below 585~keV, hence limiting the LEW range.
Consequently, the relative efficiency of prompt gamma selection has to be redefined as:
\begin{equation}
    \varepsilon^{rel, \ 124I}_{pr} = \frac{N_{pr}}{N_{pr}^{450 keV}} ,
\label{eq:eff_I}
\end{equation}
where $N_{pr}^{450 keV}$ is the number of detected prompt gammas assuming 450 keV Prompt $\gamma$ Threshold.
Additionally, the purity $\eta$ of such threshold defined as:
\begin{equation}
    \eta = \frac{N_{pr}}{N_{pr} + N_{ann}} ,
\label{eq:purity}
\end{equation}
was also checked (see Figure \ref{fig:124I}). $N_{pr}$ and $N_{ann}$ correspond to amount of registered prompt and annihilation gammas respectively, with given Prompt $\gamma$ Threshold. With rapidly decreasing $\varepsilon^{rel, \ 124I}_{pr}$, the inspected $\eta$ well resembles the overlap between annihilation and prompt gammas’ photopeaks with sigmoid-like growth saturating above 550 keV. 
Taking into account the purity $\eta$, the $FoM$ definition is extended to:
\begin{equation}
    FoM^{124I} = P_{triple} \times P_{t} \times \varepsilon^{rel}_{pr} \times \eta ,
\label{eq:fom_I}
\end{equation}

Results for $P_{triple}$, $P_{t}$, $\eta$ and $\varepsilon^{rel, \ 124I}_{pr}$ are shown in Figure \ref{fig:124I}, while the resulting figure of merit is presented in Figure \ref{fig:fom}. Assuming possible variation to the LEW range, the best result can be obtained with LEW ending at 550 keV (being 45 keV lower than the Quadra threshold) and HEW starting right after.

\subsection{Impact of cosmic and internal radiation on sensitivity}
The LSO crystals possess intrinsic radiation originating from the lutetium $^{176}$Lu isotope. $^{176}$Lu decays via the $\beta^{-}$ with a further cascade of photons with energies: 307, 202 and 88 keV \cite{Teimoorisichani2022}, which can then interact within crystals leading to the creation of false coincidences. Moreover, cosmic radiation can also be registered by the tomograph. This background was added on top of the simulation output from GATE. Based on the background data gathered with the empty Biograph Vision Quadra in Inselspital in Bern, Switzerland, the time and position distribution of the background was investigated. Then, using these distributions, the background acquisition was separately simulated and transposed in post-processing onto the simulated data. Furthermore, the current acquisition in Quadra system has an upper limit on the registered energy of about 950 keV. At the same time, the energy spectrum ends at about 726 keV, where the last bin integrates all higher depositions \cite{Will2024}. These limitations were also imposed on the corrected simulation. The resulting organ-wise sensitivities are presented in Table~\ref{tab:sensitivity}.

\begin{figure}[h!]
    \centering
    \includegraphics[width=0.7\textwidth]{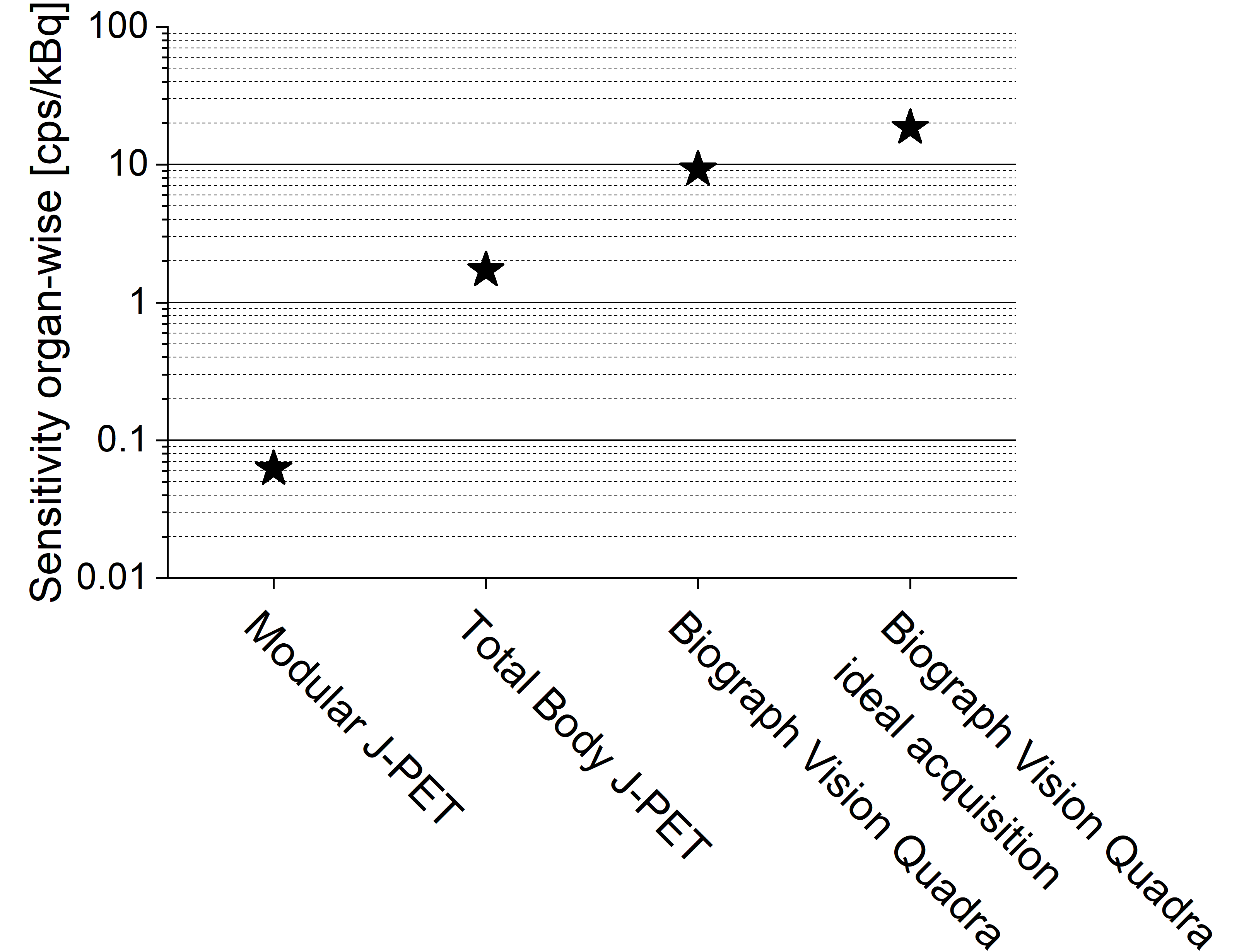}
    \caption{The organ-wise sensitivity for positronium lifetime imaging using $^{44}$Sc radioisotope estimated for the J-PET tomographs and the Biograph Vision Quadra scanner. In case of modular J-PET and Total Body J-PET the LEW was situated at 200 - 370 keV. For the Biograph Vision Quadra the standard LEW was utilized (435 - 585 keV). For all of the systems, the HEW was situated adjacent to the LEW. Calculations for the Biograph Vision Quadra were performed once assuming ideal data acquisition and once taking into account background and deposited energy measurement limitations \cite{Will2024}.}
    \label{fig:44Sc}
\end{figure}

In case of J-PET tomographs: modular J-PET and Total Body J-PET, the conditions presented in \cite{Parzych2023} were simulated. 
Organ-wise sensitivity analysis was limited to $^{44}$Sc and resulted in 0.062(08)~cps/kBq for the modular J-PET and 1.714(40)~cps/kBq for the Total Body J-PET. The mentioned uncertainties are statistical and derived from an assumption of Poisson distribution of counts in each slice based on the propagation of uncertainty. Comparison of the sensitivities obtained with $^{44}$Sc for all studied scanners is shown in Figure \ref{fig:44Sc}.

\begin{table}[h!]
 \caption{Simulated organ-wise sensitivities in cps/kBq of positronium lifetime imaging with Biograph Vision Quadra for $^{44}$Sc, $^{68}$Ga, $^{22}$Na and $^{124}$I radioisotopes
 The organ-wise sensitivity is defined as a mean of sensitivities of slices within the $\pm$4.5~cm region around the scanner’s center. The LEW was kept at standard level of 435 keV to 585 keV, while HEW was adjacent to it (Prompt $\gamma$ Threshold at 585 keV). The upper row presents results for ideal Biograph Vision Quadra data acquisition and the middle one with addition of background, upper limit at 950 keV and last bin at about 726 keV \cite{Will2024}. 
The last row takes also into account the branching ratio of the $\beta^{+}$ decay leading to the emission of prompt gamma equal to 94.3\%, 1.2\%, 90.3\% and 11.7\% for $^{44}$Sc, $^{68}$Ga, $^{22}$Na and $^{124}$I respectively \cite{Ga_decay,Sc_decay,Na_decay,I_decay}.
The presented uncertainties are statistical, and were obtained assuming Poisson distribution of counts in each slice.}
  \normalsize
\begin{tabular}{|c|c|c|c|c|}
\hline
   &
  $^{44}$Sc &
  $^{68}$Ga &
  $^{22}$Na &
  $^{124}$I \\ \hline
\begin{tabular}[c]{@{}c@{}}Biograph Vision Quadra\\ ideal acquisition\end{tabular} &
  18.53(5) &
  19.23(5) &
  18.36(5) &
  15.78(4) \\ \hline
\begin{tabular}[c]{@{}c@{}}Biograph Vision Quadra\\ current acquisition\end{tabular} &
  9.22(3) &
  10.46(4) &
  5.91(3) &
  15.39(4) \\ \hline
\begin{tabular}[c]{@{}c@{}}Biograph Vision Quadra\\ effective sensitivity\end{tabular} &
  8.70(3) &
  \begin{tabular}[c]{@{}c@{}}12.55(5)\\ $\times$ 10$^{-2}$\end{tabular} &
  5.33(3) &
  \begin{tabular}[c]{@{}c@{}}18.01(5)\\ $\times$ 10$^{-1}$\end{tabular} \\ \hline
\end{tabular}
 \label{tab:sensitivity}
\end{table}

\section{Discussion}
This simulation-based research investigated the PLI sensitivity of Quadra in an effort to optimize the energy windows criterion. The study for radioisotopes $^{68}$Ga, $^{44}$Sc, $^{22}$Na and $^{124}$I, was divided into two separate cases, due to the mutual positioning of annihilation and deexcitation photopeaks. In the case of $^{68}$Ga, $^{44}$Sc and $^{22}$Na isotopes the prompt photopeak is well isolated. This allowed for study of only Prompt $\gamma$ Threshold, while keeping the standard LEW. As shown in Figure \ref{fig:triple}, the probability of registering a true candidate for triple coincidence is almost independent of the positioning of HEW. Despite better performance by up to 6 percents to detect true coincidences among the correct candidate triples (see Figure~\ref{fig:true}), this benefit is insignificant compared to the deterioration of prompt gamma registration efficiency (see Figure~\ref{fig:eff}). As shown by means of the introduced FoM (see Figure~\ref{fig:fom}) the relative efficiency dictates the dependence on Prompt $\gamma$ Threshold. Results indicate the best performance when positioning the HEW right after LEW at 585 keV.

Due to overlapping photopeaks in the energy spectrum of the $^{124}$I radioisotope, this part of the study was conducted separately. Despite a well defined energy window for annihilation photons, its upper limit was varied in order to investigate a possible improvements in performance. In addition to all characteristics inspected previously, the purity of the prompt gamma registration in the HEW was also examined. As shown in Figure~\ref{fig:124I}, purity, probability of true candidate and true coincidence play vital roles when manipulating the LEW. Contrary, the efficiency of the deexcitation detection dominates the higher Prompt $\gamma$ Threshold regions, while other factors reach a plateau. Based on the FoM from equation \ref{eq:fom_I} (see Figure~\ref{fig:fom}), results for the HEW adjacent to the standard LEW are 20\% lower then in the case of the boundary set to 550 keV.

The introduced FoM examines the impact of sensitivity for PLI. Although higher statistics has usually a positive effect on the quality of the image, further investigations, which take into account the image reconstruction are necessary to qualitatively and quantitatively address the overall impact on PLI. 

A calculation of organ-wise sensitivities for Quadra (see Table \ref{tab:sensitivity}) was performed in two different ways: assuming an ideal data acquisition and current limitations. While for the ideal situation the best results are achieved for $^{68}$Ga, $^{44}$Sc and $^{22}$Na radioisotopes, introduction of acquisition limitations and background leads to a~significant reduction (by a factor of 2-3). This is caused by the inability of photopeak registration due to hardware limitations. However, this is not an issue for $^{124}$I with photopeak at 603 keV. Results show only slight alteration to the obtained sensitivity making the iodine the best performing radioisotope for the current Quadra system.

Investigations of PLI with $^{44}$Sc demonstrated that the organ-wise sensitivity of Quadra is 150 higher compared to the sensitivity of modular J-PET prototype.
The organ-wise sensitivity of PLI with the Total Body J-PET is expected to be 30 times higher \cite{Moskal2024} compared to the modular J-PET prototype. Regarding the sensitivity for the whole body imaging, the Total Body J-PET with a length of 250 cm and Quadra with 106 cm length will both exceed sensitivity for the whole-body positronium imaging with respect to the 50 cm length modular J-PET prototype by a factor of about 150.

\section{Conclusion}
The best performance suggest to place the lower edge of the HEW (Prompt~$\gamma$~Threshold) right after the LEW.
Moreover, to optimize the sensitivity of $^{124}$I, 
we suggest to move the upper edge of the $^{124}$I LEW down to 550 keV. 
The performed simulations indicate that the whole-body sensitivity for PLI will be more than one order of magnitude higher for Total Body J-PET and two orders for Biograph Vision Quadra, with respect to the modular J-PET prototype used to demonstrate first in-vivo poitronium lifetime images \cite{Moskal2024}.


\begin{backmatter}

\section*{Author's contributions}
The studies were conceived by P.M. and S.P. All authors contributed to the discussions, interpretation and validation of results.
Simulation of utilized systems and analysis methods were mainly developed by S.P..
P.M. and E.Ł.S. managed the whole project and secured the main financing. 
The early versions of results were interpreted mainly by S.P., S.N., and P.M.
Reconstruction of the raw experimental data were performed by S.P., S.N. and W.M.S..
The manuscript was prepared by S.P., S.N., and P.M. and was then edited and approved by all authors.

\section*{Acknowledgements}
The authors would like to thank the staff of the Inselspital, Bern University Hospital who supported the background measurement utilized in this manuscript.

\section*{Funding}
We acknowledge support from the National Science Centre of Poland through Grants No. 2021/42/A/ST2/00423, No. 2021/43/B/ST2/02150, and No. 2022/47/I/NZ7/03112. The SciMat and qLife Priority Research Area budget under the auspices of the program Excellence Initiative—Research University at Jagiellonian University. The Research Support Module as part of the Excellence Initiative – Research University program at Jagiellonian University. PLGrid (ACK Cyfronet AGH, PLG/2024/017688).

\section*{Availability of data and materials}
All data needed to evaluate the conclusions in the paper are present in the paper.

\section*{Ethics approval and consent to participate}
Not applicable

\section*{Competing interests}
P.M. is an inventor on a patent related to this work [patent nos.:
(Poland) PL 227658, (Europe) EP 3039453, and (United States) US 9,851,456], filed (Poland) 30 August 2013, (Europe) 29 August 2014, and (United States) 29 August 2014; published (Poland) 23 January 2018, (Europe) 29 April 2020, and (United States) 26 December 2017. 
W.M.S, and M.C. are employees of Siemens Medical Solutions USA, Inc., and Siemens Healthineers International AG. 
The authors declare that they have no other competing interests

\section*{Consent for publication}
Not applicable


\bibliographystyle{bmc-mathphys} 
\bibliography{bmc_article}      







\end{backmatter}
\end{document}